\documentstyle[aps,prd,preprint,epsf,floats]{revtex}
\tighten

\newcommand{\rt}{\rightarrow}

\newcommand{\ppll}{\psi^{'} \rightarrow
 \pi^+ \pi^- J/\psi$, $J/\psi \rightarrow l^+ l^-}

\newcommand{\ppee}{\psi^{'} \rightarrow
 \pi^+ \pi^- J/\psi$, $J/\psi \rightarrow e^+ e^-}

\preprint{\vbox{\hbox{BIHEP-EP1-98-02\hfill}
                \hbox{UH511-904-98\hfill}}}

\title{\boldmath Determination of $J/\psi$ Leptonic Branching Fraction
via $\psi(2S)\rightarrow \pi^{+}\pi^{-}J/\psi$}
\author{
J.~Z.~Bai$^{1}$,   J.~G.~Bian$^{1}$,  
I.~Blum$^{11}$, 
Z.~W.~Chai$^{1}$,  G.~P.~Chen$^{1}$,
H.~F.~Chen$^{10}$, 
J.~Chen$^3$,
J.~C.~Chen$^{1}$,  Y.~Chen$^{1}$,    Y.~B.~Chen$^{1}$,
Y.~Q.~Chen$^{1}$,  B.~S.~Cheng$^{1}$  X.~Z.~Cui$^{1}$,   H.~L.~Ding$^{1}$,
L.~Y.~Ding$^{1}$,  L.~Y.~Dong$^{1}$,  Z.~Z.~Du$^{1}$,    
W.~Dunwoodie$^7$,
S.~Feng$^{1}$, 
C.~S.~Gao$^{1}$,
M.~L.~Gao$^{1}$,   S.~Q.~Gao$^{1}$,   
P.~Gratton$^{11}$,
J.~H.~Gu$^{1}$,    S.~D.~Gu$^{1}$,
W.~X.~Gu$^{1}$,    Y.~F.~Gu$^{1}$,    Y.~N.~Guo$^{1}$,   S.~W.~Han$^{1}$,
Y.~Han$^{1}$,      
F.~A.~Harris$^8$,
J.~He$^{1}$,       J.~T.~He$^{1}$,    M.~He$^{5}$,
D.~G.~Hitlin$^2$,
G.~Y.~Hu$^{1}$,    H.~M.~Hu$^{1}$,    J.~L.~Hu$^{1}$,    Q.~H.~Hu$^{1}$,
T.~Hu$^{1}$,       X.~Q.~Hu$^{1}$,    J.~D.~Huang$^{1}$, Y.~Z.~Huang$^{1}$,
J.~M.~Izen$^{11}$,
C.~H.~Jiang$^{1}$, Y.~Jin$^{1}$,      Z.~J.~Ke$^{1}$,    
M.~H.~Kelsey$^2$,  B.~K.~Kim$^{11}$,  D.~Kong$^8$,
Y.~F.~Lai$^{1}$,
P.~F.~Lang$^{1}$,  
A.~Lankford$^{9}$,
C.~G.~Li$^{1}$,    D.~Li$^{1}$,       H.~B.~Li$^{1}$,
J.~Li$^{1}$,       P.~Q.~Li$^{1}$,    R.~B.~Li$^{1}$,    W.~Li$^{1}$,
W.~D.~Li$^{1}$,    W.~G.~Li$^{1}$,    X.~H.~Li$^{1}$,    X.~N.~Li$^{1}$,
H.~M.~Liu$^{1}$,   J.~Liu$^{1}$,      J.~H.~Liu$^{1}$,   R.~G.~Liu$^{1}$,
Y.~Liu$^{1}$,      
X.~C.~Lou$^{11}$,  B.~Lowery$^{11}$,
F.~Lu$^{1}$,       J.~G.~Lu$^{1}$,    J.~Y.~Lu$^{1}$,
L.~C.~Lu$^{1}$,    C.~H.~Luo$^{1}$,   A.~M.~Ma$^{1}$,    E.~C.~Ma$^{1}$,
J.~M.~Ma$^{1}$,    
R.~Malchow$^3$,
H.~S.~Mao$^{1}$,   Z.~P.~Mao$^{1}$,   X.~C.~Meng$^{1}$,
J.~Nie$^{1}$,      
S.~L.~Olsen$^8$,   J.~Oyang$^2$,      D.~Paluselli$^8$, L.~J.~Pan$^8$, 
J.~Panetta$^2$,    F.~Porter$^2$,
N.~D.~Qi$^{1}$,    X.~R.~Qi$^{1}$,    C.~D.~Qian$^{6}$,
J.~F.~Qiu$^{1}$,   Y.~H.~Qu$^{1}$,    Y.~K.~Que$^{1}$,   G.~Rong$^{1}$,
M.~Schernau$^9$,
Y.~Y.~Shao$^{1}$,  B.~W.~Shen$^{1}$,  D.~L.~Shen$^{1}$,  H.~Shen$^{1}$,
X.~Y.~Shen$^{1}$,  H.~Y.~Sheng$^{1}$, H.~Z.~Shi$^{1}$,   X.~F.~Song$^{1}$,
J.~Standifird$^{11}$,
F.~Sun$^{1}$,      H.~S.~Sun$^{1}$,   S.~Q.~Tang$^{1}$,  
W.~Toki$^3$,
G.~L.~Tong$^{1}$,
F.~Wang$^{1}$,     L.~S.~Wang$^{1}$,  L.~Z.~Wang$^{1}$,  M.~Wang$^{1}$,
Meng~Wang$^{1}$,   P.~Wang$^{1}$,     P.~L.~Wang$^{1}$,  S.~M.~Wang$^{1}$,
T.~J.~Wang$^{1}$\cite{atNU0},  Y.~Y.~Wang$^{1}$,  
M.~Weaver$^2$,
C.~L.~Wei$^{1}$,   Y.~G.~Wu$^{1}$,
D.~M.~Xi$^{1}$,    X.~M.~Xia$^{1}$,   P.~P.~Xie$^{1}$,   Y.~Xie$^{1}$,
Y.~H.~Xie$^{1}$,   W.~J.~Xiong$^{1}$, C.~C.~Xu$^{1}$,    G.~F.~Xu$^{1}$,
S.~T.~Xue$^{1}$,   J.~Yan$^{1}$,      W.~G.~Yan$^{1}$,
C.~M.~Yang$^{1}$,  C.~Y.~Yang$^{1}$,  J.~Yang$^{1}$,     
W.~Yang$^3$,
X.~F.~Yang$^{1}$,
M.~H.~Ye$^{1}$,    S.~W.~Ye$^{10}$,    Y.~X.~Ye$^{10}$,    K.~Yi~$^{1}$,
C.~S.~Yu$^{1}$,    C.~X.~Yu$^{1}$,    Y.~H.~Yu$^{4}$,   Z.~Q.~Yu$^{1}$,
Z.~T.~Yu$^{1}$,    C.~Z.~Yuan$^{1}$,  Y.~Yuan$^{1}$,     B.~Y.~Zhang$^{1}$,
C.~C.~Zhang$^{1}$, D.~H.~Zhang$^{1}$, Dehong ~Zhang$^{1}$,
H.~L.~Zhang$^{1}$, J.~Zhang$^{1}$,    J.~L.~Zhang$^{1}$, J.~W.~Zhang$^{1}$,
L.~S.~Zhang$^{1}$, Q.~J.~Zhang$^{1}$, S.~Q.~Zhang$^{1}$, X.~Y.~Zhang$^{5}$,
Y.~Zhang$^{1}$,    Y.~Y.~Zhang$^{1}$, D.~X.~Zhao$^{1}$,  
H.~W.~Zhao$^{1}$\cite{atNU1},
J.~W.~Zhao$^{1}$,  M.~Zhao$^{1}$,    W.~R.~Zhao$^{1}$, Z.~G.~Zhao$^{1}$,
 J.~P.~Zheng$^{1}$,
L.~S.~Zheng$^{1}$, Z.~P.~Zheng$^{1}$, G.~P.~Zhou$^{1}$,  H.~S.~Zhou$^{1}$,
L.~Zhou$^{1}$,     Q.~M.~Zhu$^{1}$,   Y.~C.~Zhu$^{1}$,   Y.~S.~Zhu$^{1}$,
B.~A.~Zhuang$^{1}$
\\ (BES Collaboration)}

\address{
$^1$Institute of High Energy Physics, Beijing 100039, People's Republic of
 China\\
$^2$California Institute of Technology, Pasadena, California 91125\\
$^3$Colorado State University, Fort Collins, Colorado 80523\\
$^4$Hangzhou University, Hangzhou 310028,
People's Republic of China\\
$^5$Shandong University, Jinan 250100, People's Republic of
 China\\
$^6$Shanghai Jiaotong University, Shanghai 200030,
People's Republic of China\\
$^7$Stanford Linear Accelerator Center, Stanford, California 94309\\
$^8$University of Hawaii, Honolulu, Hawaii 96822\\
$^9$University of California at Irvine, Irvine, California 92717\\
$^{10}$University of Science and Technology of China, Hefei 230026,
People's Republic of China\\
$^{11}$University of Texas at Dallas, Richardson, Texas 75083-0688}

\begin{document}
\maketitle

\begin{abstract}
A comparison of the rates for $\psi(2S)\rightarrow
\pi^+\pi^-J/\psi$, $J/\psi\rightarrow l^+l^-$ and 
$J/\psi\rightarrow$ anything is used 
to determine the $J/\psi$ leptonic 
branching fractions. 
The results are 
$B(J/\psi \rightarrow e^{+}e^{-})=(5.90\pm0.05\pm0.10)\%$ and 
$B(J/\psi \rightarrow \mu^{+}\mu^{-})=(5.84\pm0.06\pm0.10)\%$,
where the first error is 
statistical and the second is systematic. 
Assuming lepton universality, the leptonic branching 
fraction of the $J/\psi$ is 
$B(J/\psi\rightarrow \ell^{+}\ell^{-})=(5.87\pm0.04\pm0.09)\%$ per species.
This result is used to estimate the QCD scale factor $\Lambda^{(4)}_{
\overline {MS}}$ and the strong coupling 
constant $\alpha_{s}$.
\end{abstract}

\vspace*{0.3cm}


\section {Introduction}

The branching fractions for the leptonic
decays $J/\psi\rightarrow e^+e^-$ ($B_e$) and $\mu^+\mu^-$
($B_\mu$) are basic parameters of the $J/\psi$ resonance. 
They can be used to
determine the strong coupling constant $\alpha_{s}$ or, equivalently, the 
fundamental scale parameter of QCD, 
$\Lambda_{\overline{MS}}$~\cite{kwong}.
The ratio $B_e/B_\mu$ provides a test of lepton universality. 
In addition, these branching fractions are used
to determine the total number of $J/\psi$ events in a wide
variety of measurements that take advantage of the clean
experimental $J/\psi\rightarrow\ell^+\ell^-$ $(\ell=e$ or $\mu)$
signature.

The first reported measurement of the leptonic $J/\psi$ branching
fractions has a precision of about 15\% and is
based on an energy scan across the resonance performed
by the Mark-I group~\cite{mk1}.   A subsequent measurement
by the Mark-III group~\cite{mk3} is
based on a comparison of the rates for $\psi(2S)\rightarrow
\pi^+\pi^-J/\psi$, $J/\psi\rightarrow \ell^+\ell^-$ and 
$J/\psi\rightarrow$ anything and is, thus,  
independent of the luminosity determination.
The Mark-III measurement has a precision of 4\% 
and is about one (Mark I) standard deviation below the Mark-I result. 
More recently, BES performed an energy scan measurement and obtained
results in good agreement with Mark-III, but with larger errors ~\cite{zhuys}. 
In this paper, we report the results of a measurement
of the leptonic branching fractions using a sample of
$\psi(2S)$ decays measured in the BES detector at the
BEPC storage ring.  We apply a technique similar to that
used by the Mark-III group to a larger data sample. 

\section{The BES detector}

The Beijing Electron Spectrometer, BES,
is a conventional cylindrical magnetic detector that is coaxial
with the BEPC colliding $e^+e^-$ beams.  It is 
described in detail in Ref.~\cite{bes}. A four-layer central drift
chamber (CDC) surrounding the beampipe provides trigger
information. Outside the CDC, the forty-layer main drift chamber (MDC)
provides tracking and energy-loss ($dE/dx$) information on
charged tracks over $85\%$ of the total solid angle.
The momentum resolution is $\sigma _p/p = 0.017 \sqrt{1+p^2}$ ($p$
in GeV/c), and the $dE/dx$ resolution for hadron tracks for this
data sample is $\sim 9\%$. 
An array of 48 scintillation counters surrounding the MDC provides 
measurements of the time-of-flight (TOF) of charged tracks with a resolution of
$\sim 450$ ps for hadrons. Outside the TOF system, a 12
radiation length lead-gas barrel shower counter (BSC),
operating in self-quenching streamer mode, measures the energies 
of electrons and photons over  80\% of the total solid
angle. The energy resolution is $\sigma_E/E=0.22/\sqrt{E}$ ($E$
in GeV).
Surrounding the BSC is a solenoidal magnet that
provides a 0.4 Tesla magnetic field in the central tracking
region of the detector. Three double layers of proportional chambers
instrument the magnet flux return (MUID) and are used to identify
muons of momentum greater than 0.5 GeV/c. Endcap time-of-flight
and shower counters extend coverage to the forward and backward
regions.

\section{Technique}

Our measurement is based on a data sample corresponding to
an integrated luminosity of about 6.1 pb$^{-1}$ 
accumulated at the $\psi(2S)$ resonance.  The 
$\psi(2S)$ is a copious source of $J/\psi$ decays: 
the branching fraction 
$\psi(2S)\rightarrow \pi^+\pi^- J/\psi = 0.324\pm 0.026$~\cite{pdg},
is the largest single $\psi(2S)$ decay channel.
We determine the $J/\psi$ leptonic branching
fraction from a comparison of the exclusive and inclusive processes:
\begin{center}
\begin{tabbing} 
\hspace{20 mm}\= $\psi(2S) \rightarrow \pi^+\pi^-$ $J$\=$/\psi$           \\
              \>         \> $\hookrightarrow l^+l^- $\hspace{20 mm}\= $(I)$ \\
  and         \>         \> $\hookrightarrow$  anything \>$(II)$      \\
\end{tabbing}
\end{center}
The $J/\psi$ leptonic branching fraction is determined from the relation
\begin{eqnarray*}
B(J/\psi\rightarrow l^{+}l^{-})&=&
\frac{N^{obs}_{\ell}/\varepsilon_{\ell}}
          {N^{obs}_{J/\psi}/\varepsilon_{J/\psi}},
\end{eqnarray*}
where $N^{obs}_{\ell}$ and $N^{obs}_{J/\psi}$ are observed numbers of events
for processes I and II, and
$\varepsilon_\ell$ and $\varepsilon_{J/\psi}$ are the respective acceptances.

\section{Event Selection}

For both processes I and II, we use only runs of good quality. We
require at least one pair
of oppositely charged candidate pion tracks that each
satisfy the following criteria:
\begin{enumerate}
\item   $P_{\pi}<0.5$ GeV/c,
where $P_{\pi}$ is the pion momentum. 

\item   $P_{\pi xy}>0.1$ GeV/c,
where $P_{\pi xy}$ is the momentum of the pion transverse to the beam
direction.  This removes tracks that circle in the Main Drift Chamber.


\item   $|\cos\theta_{\pi}|<0.75$.
Here $\theta_{\pi}$ is the polar angle of the $\pi$ in the laboratory system.

\item $\cos \theta_{\pi \pi} < 0.9 $.
$\theta_{\pi \pi}$ is the laboratory angle between the $\pi^+$ and $\pi^-$.
This cut is used to eliminate contamination from misidentified
$e^+e^-$ pairs from $\gamma$ conversions, as shown in Fig.~\ref{fig:cospipi}.

\begin{figure}[!htb]
\centerline{\epsfysize 3.0 truein
\epsfbox{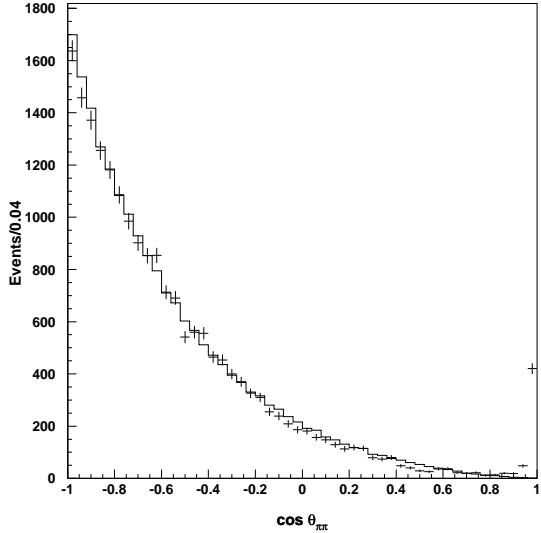}}
\caption{\label{fig:cospipi} The distribution of the cosine of the angle 
between the $\pi^+$ and $\pi^-$, $\cos \theta_{\pi \pi}$, 
in
$\ppee$ events (dots with error bars) compared with the distribution
of Monte Carlo data (histogram).
The data has background near $\cos \theta_{\pi \pi} = 1$.
}
\end{figure}

\end{enumerate}

\noindent
The invariant mass recoiling against the candidate $\pi^+\pi^-$
pair, $m^{recoil}_{\pi^+\pi^-} = [(m_{\psi(2S)} - E_{\pi^+} - E_{\pi^-})^2
 - ({\bf p}_{\pi^+} + {\bf p}_{\pi^-})^2]^{\frac{1}{2}} $, is required to be in 
the range $3.0\le m^{recoil}_{\pi^+\pi^-}\le 3.2~$GeV/c$^2.$

\subsection{\boldmath $J/\psi\rightarrow\ell^+\ell^-$}
For leptonic decay candidate events (process-I),
the number of charged tracks is required to be at least four
with a 4-track combination of net charge zero \cite{4prng}. Lepton pair candidates must
satisfy the following selection criteria:

\begin{enumerate}
\item   
$P_{l} >0.5$ GeV/c.
Here $P_l$ is the three-momenta of
the candidate lepton track.

\item   
$P_{l^+}>1.3$ GeV/c or $P_{l^-} >1.3$ GeV/c or
$(P_{l^+} + P_{l^-}) > 2.4$ GeV/c.  This cut
selects events consistent with $J/\psi$ decay, while rejecting
background. 

\item  
$|\cos\theta_{e}|<0.75$, $|\cos\theta_{\mu}|<0.60$.
Here $\theta_{e}$ and $\theta_{\mu}$ are the polar angles of the
electron and muon, respectively.
This cut ensures that 
electrons are contained in the $BSC$ and muons in
the MUID system.

\item
$\cos \theta^{cm}_{l^+ l^-} < -0.975$,
where $ \theta^{cm}_{l^+ l^-}$ is the angle between the two leptons in
the rest frame of the $J/\psi$.

\item 
For $e^+e^-$ candidate pairs: $SCE_+$ and $SCE_- >
0.6$ GeV/c, 
where $SCE$ is the energy deposited in the BSC, or, if one of the tracks goes
through a BSC rib or has $P_l < 0.8 $ GeV/c, the $dE/dx$ information of
both tracks in the MDC must
be consistent with that expected for electrons. The rib region of the
BSC is not used because the Monte Carlo does not model the energy deposition
well in this region. 

\item 
For $\mu^{+}\mu^{-}$ pair candidates at least one track must have
$N^{hit} > 1$,
where $N^{hit}$ is the number of MUID layers with matched hits 
and ranges from 0 to 3. If only one track is identified in this fashion, then
the invariant mass of the $\mu \mu$ pair must also be within 250 MeV/c$^2$ of
the $J/\psi$ mass. 

\end{enumerate}

Fig.~2a shows the $m^{recoil}_{\pi^+\pi^-}$
distribution for the $\ppll$ events.

\begin{figure}[!htb]
\centerline{\epsfysize 3.5 truein
\epsfbox{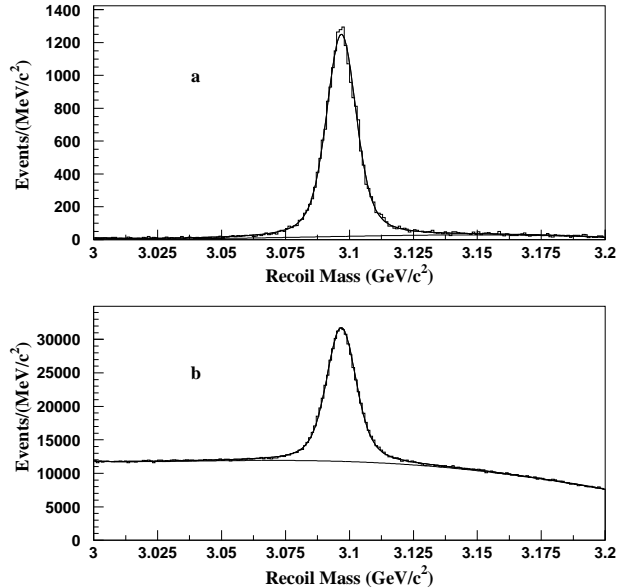}}
\caption{\label{fig:npsip} {\bf a)} Number of events versus
$m^{recoil}_{\pi^+\pi^-}$, the mass recoiling against the two $\pi$'s, for
$\psi(2S) \rightarrow \pi^+ \pi^- J/\psi, J/\psi \rightarrow l^+ l^-$
events.  The histogram is data, and the smooth
curve is a BWG (``signal'') plus a third order polynomial (``background'').
{\bf b)} Number of events versus $m^{recoil}_{\pi^+\pi^-}$ for inclusive events. The
histogram is data, and the smooth curve is a BWG (signal) with parameters
determined from a) plus a fourth order polynomial (background).
}
\end{figure}

\subsection{Fitting}

The number of process-II events  ($J/\psi\rightarrow$ anything)
is determined from  a fit to the
$\pi^{+}\pi^{-}$ recoil mass spectrum, 
using a  $J/\psi$ line shape that
is determined
from the  recoil mass spectrum for the leptonic decays (Fig.~2a).
This is fit with a 
Breit-Wigner function folded with a Gaussian (BWG) and a third order
background polynomial~\cite{tails}.  The
parameters thus obtained for the BWG fit
are then used together
with a fourth order polynomial background to fit the $\pi^{+}\pi^{-}$ 
recoil mass for the inclusive $J/\psi$ decays, as shown in
Fig.~2b. 
The resulting  number of inclusive decays is
$N_{J}^{obs}=530423 \pm 1270$, 
where the error is the statistical uncertainty combined with the uncertainty 
in the fitting procedure.  The same BWG parameters and a third order
background are used to obtain the number of $e^+ e^-$ and $\mu^+ \mu^-$ events
separately and yield a
total of $18118 \pm 150$ $J/\psi \rt e^+ e^-$ events and $14611 \pm 134$
$J/\psi \rt \mu^+ \mu^-$ events.  The results are summarized in
Table~\ref{summary}.
\begin{table}[h]
\caption{\label{summary}Summary of Results.}
\begin{center}
\begin{tabular}{c|c|c|c}
& $J/\psi\rightarrow$ anything & $J/\psi\rightarrow e^+e^-$ &
         $J/\psi\rightarrow \mu^+\mu^-$             \\\hline
$N^{obs}$        &$530423 \pm 1270$   & $18118 \pm 150$    &$14611 \pm 134$    \\
$\epsilon^{MC}$         &$44.67\%$  & $25.85\%$ &$21.07\%$ \\\hline
$N^{obs}/\epsilon^{MC}$ &$1.1876 \times 10^6 $  & $70089$ &$69345$ \\
\end{tabular}
\end{center}
\end{table}

\section{Acceptance}

The acceptances are obtained from Monte Carlo simulations.
According to ref.~\cite{pham}, the orbital angular momenta between
the $\pi^{+}\pi^{-}$ system and the $J/\psi$,  as well as that
between the $\pi^{+}$
and $\pi^{-}$ is zero, and
the $\pi^{+}\pi^{-}$ mass, $m_X$, distribution is
\begin{eqnarray*}
\frac{d\sigma}{dm_{X}} & \propto & ({\rm Phase~Space}) \times 
(M^{2}_{X}-4m^{2}_{\pi})^{2}   \\
& \propto & (M^{2}_{X}-4m^{2}_{\pi})^{2}(M^{2}_{X}-4m^{2}_{\pi})^{1/2} \times \\
&&[(M^{2}_{\psi(2S)}-m^{2}_{J/\psi}-m^{2}_{X})^{2}
-4m^{2}_{J/\psi}m^{2}_{X}]^{1/2}.
\end{eqnarray*}

\begin{figure}[!htb]
\centerline{\epsfysize 3.0 truein
\epsfbox{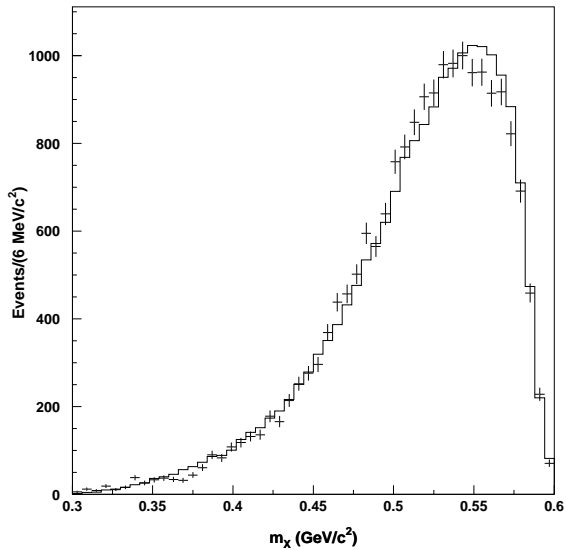}}
\caption{\label{fig:mpipi}
Number of events versus $m_X$.  The points with error bars are
data, and the histogram is Monte Carlo data.
}
\end{figure}

We generate process-I as sequential two-body decays
$\psi(2S) \rightarrow X+J/\psi$, $X\rightarrow \pi^+\pi^-$, and 
$J/\psi \rightarrow l^+l^-$.
Isotropic angular distributions are used for the $J/\psi$ in the
laboratory and for the charged pions in the $X$ rest frame~\cite{dwave}.
Leptons are generated with a  $ 1+\cos^2\theta$ angular distribution 
in the $J/\psi$ rest frame and with order $\alpha^3$ final state 
radiative corrections~\cite{berends}. 
The $\pi^+$ and $\pi^-$ decay in the detector according
to the PDG\cite{pdg}
life time and branching ratios.
Initial state radiation is not included
in the $\psi(2S)$ generation for data taken at
the $\psi(2S)$ peak energy. 
See Fig.~\ref{fig:mpipi} for a comparison of the $m_X$ distribution for
data and Monte Carlo generated data.

Monte Carlo samples of about eight times the number of events produced
in our data samples via process-I
are generated.
After application of the same selection criteria and fitting procedure as used
for the data, we get
$\varepsilon_{e}=25.85\%\ $ and
$\varepsilon_{\mu}=21.07\% $.

For process-II, the $\pi \pi$ acceptance, $\varepsilon_{J/\psi}$,
depends on the charged particle
multiplicity produced in the $J/\psi$ decay.    
The acceptances obtained from 
generating different multiplicity $J/\psi$ decay events
are listed in Table~\ref{accept}.  Also shown is the
multiplicity distribution obtained from our data\cite{mctd1}.
Using this distribution and the acceptances, we
obtain
$\varepsilon_{J/\psi}=(44.67 \pm 0.20)\%$,
where the error comes from varying the multiplicity values over the
range of values reported by other experiments\cite{mctd2}, as well as
the Monte Carlo statistical uncertainties.
The acceptance $\epsilon_{J/\psi}$ is not sensitive to the
number of $\pi^{0}$'s or $K^o$'s accompanying the charged $\pi$'s in
the $J/\psi$ decay channel or to replacing the charged $\pi$'s with charged kaons.
\begin{table}[h]
\begin{center}
\caption{\label{accept}Acceptance ($\pi \pi$) for different charged multiplicities.}
\begin{tabular} {cccc}
Charged      & MC decay ~  &~ Percent            ~&~Acceptance (\%)  \\
Multiplicity &  channel    &   of $J/\psi$~       &            \\   \hline
0   & all neutrals          &  0.5   & 48.34                  \\
2   & $\pi^+\pi^-\pi^0$      &  35.6  & 46.47                  \\ 
4   & $2(\pi^+\pi^-)$        &  40.9  & 44.26                  \\ 
6   & $3(\pi^+\pi^-)$        &  17.7  & 43.03                  \\ 
8   & $4(\pi^+\pi^-)$        &   5.3  & 41.32                  \\
\end{tabular}
\end{center}
\end{table}

\section{Error Estimation}
 
The statistical branching fraction errors, propagated from the 
statistical errors on  $N^{obs}_{l}$ and $N^{obs}_{J}$, 
are 
$\sigma_{B_{e}}=0.0005$ and
$\sigma_{B_{\mu}}=0.0006$.


The effect of changing cuts has been studied.  The results are
shown in Table~\ref{tab:varoldandnew}.
Other systematic contributions come from 
the acceptance uncertainties, 
particle identification (PID) uncertainties,
and the fitting method.
For $e e$ events, electron particle identification backgrounds
and efficiencies can be
measured by comparing the BSC results
and the $dE/dx$ information from the MDC for events outside the BSC
ribs.
The backgrounds found and the difference in efficiency between that
determined from the data and that predicted by the
Monte Carlo are small and are used in the estimate of the electron PID 
systematic error. The PID systematic error for the $\mu \mu$ events is
obtained by determining the background allowed by the $\mu \mu$ invariant
mass cut for events where only one $\mu$ track is identified by the MUID
system and the estimated efficiency of this cut.   
The systematic error associated with the fitting procedure is determined
by using an alternative fitting method.
The total systematic error is taken as the sum in quadrature
of all the individual systematic errors. The relative systematic
errors on both the $J/\psi \rightarrow e^+e^-$ 
and $\mu^+\mu^-$ branching fractions are 1.7\%. 

\begin{table}[h]
\caption{Branching Ratio Systematic Errors (\%)}
\
\begin{center}
\label{tab:varoldandnew}
\begin {tabular}{llrr}
Variable~~~~&~~Variation~~ & $B(e e)$&$ B(\mu \mu)$ \\ \hline
$|\cos \theta_{\pi}|$ cut  & $0.75 \rt 0.70$ & 0.21  & 0.74 \\
$\cos \theta_{\pi \pi}$ cut & $0.9 \rt 0.85$ & 0.07  & 0.07 \\
$P_{\pi xy}$ cut     & turn off              & 0.46  & 0.56 \\
$|\cos \theta_{\mu}|$ cut & $0.6 \rt 0.65$   & 0.05  & 0.23 \\
$|\cos \theta_e|$ cut & $0.75 \rt 0.7$       & 0.19  & 0.10 \\
$\cos \theta_{l^+ l^-}^{cm}$ cuts & turn off & 0.46  & 0.28 \\
SCE cut        &       $0.6 \rt 0.7$ GeV     & 0.18  &      \\
$P_l^+$ or $P_l^-$ cut& $> 1.3 \rt > 1.4$ GeV/c & 0.30  & 0.01 \\
$P_l^+$ and $P_l^-$ cut& $> 0.5 \rt > 0.8$ GeV/c & 0.89  & 0.09 \\
Use only best tracks   &                     & 0.14  & 0.43 \\
Fitting Method         &                     & 0.85  & 1.21 \\
Efficiency Uncertainty &                     & 0.41  & 0.45 \\
PID Uncertainties      &                     & 0.84  & 0.17 \\\hline
Combined    &  &  1.69 & 1.71 \\
\end{tabular}
\end{center}
\end{table}

Contamination from hadronic events has been checked by using a sample
of kinematically selected $\psi(2S) \rt \pi^+ \pi^- J/\psi, J/\psi \rt
\rho \pi$ events.  The number of these
events satisfying the cuts used in this analysis was negligible.  In
addition, a simulation of the potential background process $\psi(2S)
\rightarrow \eta J/\psi$ with $\eta$ going to $\pi^+\pi^-\pi^0$ or
$\pi^+\pi^-\gamma$ indicates a negligible contribution.

\section{Results and Discussion}
The numbers of events obtained are summarized in 
Table~\ref{summary} \cite{ycz},
and the final results for the branching fractions are:
$$B(J/\psi \rightarrow e^+e^-)=(5.90 \pm 0.05 \pm 0.10)\%$$
and
$$B(J/\psi \rightarrow \mu^+\mu^-)=(5.84 \pm 0.06 \pm 0.10)\%.$$
The close equality of $B_{\mu}$ and $B_{e}$ is a verification
of $e$-$\mu$ universality:
$$\frac{B(J/\psi \rightarrow e^+e^-)}{B(J/\psi \rightarrow \mu^+\mu^-)}
=1.011 \pm 0.013 \pm 0.016. $$
Assuming 
$B_{\mu}=B_{e}$ \cite{phase}, we find a combined leptonic branching fraction of
\[B(J/\psi \rightarrow \ell^+\ell^-)=(5.87 \pm 0.04 \pm 0.09)\%\]

Our results are compared with previous experiments in Table~\ref{tab:compare}.
They are consistent with and improve on the precision of
the  MARK-III measurement~\cite{mk3}.
They are also consistent with BES  
results determined from $e^+e^-$
cross section measurements in the vicinity of $J/\psi$ 
resonance~\cite{zhuys}.
If we combine the
values of $B(ll)$, we obtain a new world average
\[B(ll) = (5.894 \pm 0.086)\%, \]
which has an error about half that in the 1996 PDG\cite{pdg}.

\begin{table}[!h]
\caption{Comparison with other experiments}
\
\begin{center}
\begin {tabular}{lll}
\label{tab:compare}
Experiment~~~~~~~~~&~~~~$B(ee) (\%)$ ~~~~~&~~~$B(\mu \mu) (\%)$~ \\ \hline
Mark I~\cite{mk1}    & $6.9 \pm 0.9 $   &  $6.9 \pm 0.9$ \\
Mark III~\cite{mk3}& $5.92 \pm 0.15 \pm 0.20$   &  $5.90 \pm 0.15 \pm 0.19$\\
BES Scan~\cite{zhuys}   & $6.09 \pm 0.33$             &  $6.08 \pm 0.33$  \\
THIS EXPERIMENT        & $5.90 \pm 0.05 \pm 0.10$ & $5.84 \pm 0.06 \pm 0.10$ \\
\end{tabular}
\end{center}
\end{table}

The experimental ratio of the quarkonium annihilation rates
$$R_2 = \frac{\Gamma(quarkonium\rightarrow ggg)}
{\Gamma(quarkonium\rightarrow \mu^+\mu^-)}$$
allows the determination of the strong coupling constant
$\alpha_{s}$ or, equivalently, the QCD scale parameter 
$\Lambda^{(n_{f})}_{\overline{MS}}$~\cite{kwong,chiang}.
Following the notation of ref.~\cite{chiang}, one obtains
\begin{eqnarray*}
(1+R_{1})R_{2} & = &[1-2B(J/\psi\rightarrow l^{+}l^{-})       \\
         & - &B(J/\psi\rightarrow \gamma^{*} \rightarrow q\bar{q}) \\
         & - &B(J/\psi\rightarrow \gamma \eta_{c})]/
     B(J/\psi\rightarrow l^{+}l^{-})
\end{eqnarray*}
where,
$$ R_{1}=\frac{\Gamma(J/\psi\rightarrow \gamma gg)}
                    {\Gamma(J/\psi\rightarrow ggg)}
              =\frac{16\alpha}{5\alpha_{s}}(1-3.0\frac{\alpha_{s}}{\pi}), $$
$$ R_{2}=\frac{\Gamma(J/\psi\rightarrow ggg)}
              {\Gamma(J/\psi\rightarrow l^{+}l^{-})}
        =\frac{5(\pi^{2}-9)\alpha^{3}_{s}}
              {18\pi\alpha^{2}}
          (1+1.59\frac{\alpha_{s}}{\pi})\gamma_{\psi}, $$
and 
$$ \gamma_{\psi}=\frac{\Gamma(c\bar{c}\rightarrow ggg)}
                          {\Gamma^{0}(c\bar{c}\rightarrow ggg)}
                  =0.31 \pm 0.03.$$
The factor $\gamma_{\psi}$ is the reduction factor of the three
gluon decay caused by the finite size effect in the matrix element. 
For $R_1$, we use 
the theoretically calculated rather than the measured value because
of the large error associated with the latter~\cite{scharre}.

Combining the above $J/\psi$ decay parameters and
the values for
$B(J/\psi\rightarrow \gamma^{*} \rightarrow q\bar{q})$ and
$B(J/\psi\rightarrow \gamma \eta_{c})$ listed in PDG\cite{pdg},
we obtain 
$$\alpha_{s}(m_c=1.5GeV/c^2)=0.28 \pm 0.01~~~{\rm and }$$
$$\Lambda^{(4)}_{\overline{MS}}=(209 \pm 21) {\rm MeV,}$$
where only the 
experimental errors are included.
Our results are in good agreement with
the PDG\cite{pdg} values.


We would like to thank the staff of BEPC accelerator and the IHEP Computing
Center for their efforts.



\begin{references}
\bibitem{4prng} Approximately 4\% of the events have more than four
prongs.  These are usually due to very low energy $\pi$ track being
reconstructed into two tracks.  The Monte Carlo
does not model this effect well, so it is necessary to include these events,
even though the $m^{recoil}_{\pi^+ \pi^-}$ distribution is not as clean
for these events. We rely on our fitting procedure to extract the
signal.
\bibitem{tails} The events represented by the polynomial in Fig 2a
are predominantly good events but where the low energy pion underwent
$\pi \rt \mu \nu$ decay. Using only the BWG shape to describe the signal makes
us less sensitive to background.
\bibitem{dwave} We have also determined the acceptance allowing the
$\pi \pi$ system to have a small amount of D-wave.  This has negligible
effect on the results.
\bibitem{mctd1}
For process-II, $n^{obs}_{i}$, the total number of events observed with
charged multiplicity $i$ is related to the $N^{prod}_{k}$, the
number generated with charged multiplicity $k$, by the relation 
\[n^{obs}_{i}=\sum_{k} \varepsilon_{ki} N^{prod}_{k} \]
\centerline{with $k=0,2,4,\ldots,\;\;\; i=0,1,2,3,\ldots ,$}
where the matrix element
$\varepsilon_{ki}$ is the probability of an observed
$i$-$prong$ event coming from a $k$-$type$ event.
Summing over $i$ and using
 \[\frac{n^{obs}}{N^{prod}}=\varepsilon_{J/\psi},\] 
we have
\[\varepsilon_{J/\psi}=\sum_{k}\varepsilon_{k}\omega_{k}\]
where $\varepsilon_{k}$ is the acceptance for
and $\omega_{k}$ the fraction of multiplicity-$k$ events.
By solving the overdetermined
equations above, we estimated $J/\psi$ 
charged-track-multiplicity distribution shown in Table~\ref{accept}.
\bibitem{ycz} Using variations like those in Table~\ref{tab:varoldandnew},
we can determine the systematic errors associated with $N_{J/\psi}$.
We obtain $N_{J/\psi} = (1.188 \pm 0.003 \pm 0.016) \times 10^6$.
Using the branching
ratio of $\psi(2S) \rightarrow \pi^+\pi^-J/\psi$ given by \cite{pdg}, 
the produced number of $\psi(2S)$ events corresponding to the data
used in this analysis is estimated to be 
$(3.67  \pm 0.01 \pm 0.30) \times 10^{6}$.
Other analyses may use, in addition, runs of fair quality.
Using
good and fair quality runs, yields $(1.227 \pm 0.003 \pm 0.017) \times
10^6$ $J/\psi$
events and $(3.79 \pm 0.01 \pm 0.31) \times 10^6$ $\psi(2S)$ events.
\bibitem{phase} The difference in the branching ratios due to the difference
in phase space for the two modes is small compared to the experimental errors.
\end{references}
\end{document}